# RECOVERY OF 150-250 MeV COSMIC RAY PROTON INTENSITIES BETWEEN 2004-2010 AS MEASURED NEAR THE EARTH, AT VOYAGER 2 AND ALSO IN THE HELIOSHEATH AT VOYAGER 1 – A TWO ZONE HELIOSPHERE

**W.R Webber[1], F.B. McDonald[2], P.R. Higbie[3], B. Heikkila[4]**

1. New Mexico State University, Department of Astronomy, Las Cruces, NM, 88003, USA
2. University of Maryland, Institute of Physical Science and Technology, College Park, MD, 20742, USA
3. New Mexico State University, Physics Department, Las Cruces, NM, 88003, USA
4. NASA/Goddard Space Flight Center, Greenbelt, MD 20771, USA

**Abstract**

The recovery of cosmic ray protons of energy ~150-250 MeV/nuc in solar cycle #23 from 2004 to 2010 has been followed at the Earth using IMP, ACE and balloon data and also at V2 between 74-92 AU and at V1 beyond the heliospheric termination shock (91-113 AU). The correlation coefficient between the intensities the Earth and V1 during this time period, is 0.936, allowing for a ~0.9 year delay due to the solar wind propagation time from the Earth to the outer heliosphere. To describe these intensity changes and to predict the absolute intensities measured at all three locations we have used a simple spherically symmetric (no drift) two-zone heliospheric transport model with specific values for the diffusion coefficient in both the inner and outer zones. The diffusion coefficient in the outer zone, from about 90 to 120 (130) AU, is determined to be ~5-10 times smaller than that in the inner zone out to 90 AU. This means that the outer zone acts much like a diffusing barrier in this model. The absolute magnitude of the intensities and the intensity changes at V1 and the Earth are described to within a few percent by a diffusion coefficient that varies with time by a factor ~4 in the inner zone and only ~1.8 (1.25) in the outer zone over the time period from 2004-2010. These diffusion coefficients and their variations are essentially the same as those derived earlier from a similar study using He nuclei of the same energy. This model and the diffusion coefficients used provide a total modulation potential at the Earth ~250 MV in 2009. The difference ~10-20% between calculated and observed intensities at V2 can be explained if the heliosphere is squashed by ~10% in distance (non-spherical) so that the HTS is closer to the Sun in the direction of V2 compared to V1.

**Introduction**

The intensity recovery of galactic cosmic rays at the Earth in the current solar 11-year cycle between 2004-2009 is well documented using spacecraft data (e.g., McDonald, Webber and Reames, 2010; Mewaldt, et al., 2009, 2010). This cosmic ray recovery started in early 2004 at the Earth after the large "Halloween" events in October-November, 2003, and has been observed by neutron monitors and various spacecraft near the Earth including ACE, IMP and others. This recovery was observed by V2 and V1 to begin in the outer heliosphere in late 2004 after the Halloween event had propagated out to their respective locations at 76 and 93 AU (McDonald, et al., 2006). At the end of 2004 V1 crossed the HTS at 94 AU and has continued to move outward so that by 2010.5 it was at ~114 AU, perhaps ~30 AU or more beyond the current HTS location, estimated to be between 80-85 AU (Webber and Intriligator, 2011). Thus V1 has spent essentially the entire recovery cycle beyond the HTS in the heliosheath region where the solar wind parameters are measurably different from those in the inner heliosphere. V2 remained in the "inner" part of the heliosphere until 2007.66 when at ~84 AU it also crossed the HTS.

At about 2010.0 the cosmic ray proton intensity at the Earth reached its maximum. At V1 the intensity continues to increase as of 2010.5 whereas at V2 it reached a maximum in early 2009. At the Earth the proton intensities reached levels ~25% higher than those observed during the previous 11-year intensity maximum in 1997-98 (McDonald, Webber and Reames, 2010; Mewaldt, et al., 2009, 2010). At V1 the cosmic ray proton intensities are at the highest levels yet observed in the heliosphere and at energies ~200 MeV/nuc at 2010.5 are within ~20% of the estimated LIS intensities for protons (see Webber and Higbie, 2009).

It is the purpose of this paper to compare the proton intensities between 150-250 MeV/nuc observed at the Earth and those observed at V1 and V2 during this time period within the framework of a simple modulation model, with the objective of understanding better the global characteristics of the solar 11-year modulation cycle, including particularly the modulation effects beyond the HTS in the heliosheath.

This is the second of several articles dealing with the recovery of cosmic ray intensities at V1, V2 and the Earth during this extended time period. The first article considered the 150-250 MeV He nuclei (Webber, et al., 2011). A following article will study the 20-125 MeV/nuc Carbon nuclei. Each particle provides its own specific information on the heliosphere modulation process and the required "source" spectrum of the particles involved.

**Observations at the Earth and at V1 and V2**

In Figure 1 we show the time history of ~130-250 MeV/nuc protons at the Earth from 1998 to the present time. This data is smoothed by taking 5 times 26 day moving averages. The data at the Earth is IMP-8 data from 1998 to 2006 from McDonald and Heikilla (2009, private communications), and ACE data from 2006-2010. The ACE data is for E >120 MeV protons from 1997-2010 as presented by Mewaldt, et al., 2010. The correlation between the two data sets is excellent between 1998-2006. The regression line between data sets has a logarithmic slope of $2.05 \pm 0.05$ and this is used to extend the IMP-8 data from 2006 to 2010.

Also shown in this Figure is the corresponding data for ~155-245 MeV/nuc protons at V1 and V2 corrected for a background (10-20%) of low energy ACR H. At the beginning of the recovery time period the intensity at V1 was ~8 times that at the Earth. This is a measure of the overall interplanetary gradient between 1 and ~94 AU, the location of V1 at that time. By 2010.5 this intensity ratio is reduced to ~4 implying that the intensity changes between 2004 and 2010.5 at the Earth are greater than those at V1. This changing intensity ratio at V1 and also at V2 is shown in Figure 2.

In Figure 3 we show the proton data at the Earth from 2004-2010 superimposed on the data at V1 (with different intensity scales), with the data at Earth delayed to account for the solar wind propagation time from the Earth to V1. This delay time is varied from 0.5 to 1.5 years in 26 day increments and the correlation coefficient reaches a maximum value of 0.936 for time delays between 0.86 and 0.93 years. This correspondence of time histories is remarkable considering the 100 AU difference in the radial location of the spacecraft.

This correlation throughout the heliosphere is also evident in Figure 4A which shows the intensities at V1 and V2 vs. those at the Earth, with a delay ~0.89 yrs. The "loop" in the regression curves in Figure 4A between Earth data and the V1 and V2 data is due to the largest transient cosmic ray decrease in solar cycle #23 (the September, 2005, event at the Earth) propagating outward through the heliosphere, reaching V2 at ~2006.15 and V1 at about 2006.5. If this time period is excluded from the correlation calculation, the maximum value for the correlation coefficient between V1 and the Earth intensities increases to 0.968 for a delay of 0.89 years.

We seek to fit the data in Figure 4A and to interpret it using a simple global modulation model. This model should predict the absolute intensities at all three locations and also the changing ratios of intensities at V1 beyond the HTS, and at V2 mainly just inside the HTS, to those intensities at the Earth vs. time as given by Figure 2 and also Figure 4A as well as the slope of the regression lines between V1 and V2 and the Earth, that is the ratio of the rates of change of intensity at each location as shown in Figure 4A.

From Figure 1 we observe that the proton intensities at V2 were slightly less than those at V1 during the minimum modulation period from 1998 to the middle of 2000. Then with increased modulation a larger radial gradient was established between the two spacecraft which continued until after the large transient decrease in 2006 noted above passed V2 and then V1. From mid 2007 to mid 2009 the intensity at V2 (delayed by 0.21 year, the transit time of a solar disturbance between V2 and V1) was again slightly less than the intensity at V1. Then in 2009 the intensity at V2 stopped increasing and by 2010 the difference in V1 and V2 intensities was ~20% implying again a larger radial gradient between the two spacecraft. These intensity differences between V1 and V2 for protons during this time period are significantly different than those observed for the same energy He nuclei (see Webber, et al., 2011).

## The Cosmic Ray Transport Equation in the Heliosphere

Here we use a simple spherically symmetric quasisteady state no-drift transport model for cosmic rays in the heliosphere. While this simplified model obviously cannot fit all types of observations it does provide a useful insight into the inner heliospheric/outer heliospheric modulation and helps to determine which aspects of this modulation need more sophisticated models for their explanation. The numerical model was originally provided to us by Moraal (2003) and is similar to the model described originally in Reinecke, Moraal and McDonald, 1993, and in Caballero-Lopez and Moraal (2004), and also to the spherically symmetric transport model described by Jokipii, Kota and Merenyi, 1993 (Figure 3 of that paper). The basic transport equation is (Gleeson and Urch, 1971);

$$\frac{\partial f}{\partial t} + \nabla \cdot (CVf - K \cdot \nabla f) + \frac{1}{3p^2} \frac{\partial}{\partial p} (p^2 V \cdot \nabla f) - Q$$

Here f is the cosmic ray distribution function, p is momentum, V is the solar wind velocity, K(r,p,t) is the diffusion tensor, Q is a source term and C is the so called Compton-Getting coefficient.

For spherical symmetry (and considering latitude effects to be unimportant for this calculation) the diffusion tensor becomes a single radial coefficient $K_{rr}$. We assume that this coefficient is separable in the form $K_r(r,P) = \beta K_1(P) K_2(r)$, where the rigidity part, $K_1(P) \equiv K1$ and radial part, $K_2(r) \equiv K2$. The rigidity dependence of $K(P)$ is assumed to be $\sim P$ above a low rigidity limit $P_B$. The units of the coefficient $K_{rr}$ are in terms of the solar wind speed $V = 4 \cdot 10^2$ km·s$^{-1}$, 1 AU = 1.5 x $10^8$ km, so $K_{rr} = 6 \cdot 10^{20}$ cm$^2$·s$^{-1}$ when K1 = 1.0.

We consider two possible scenarios. The first is a simple heliosphere with the diffusion coefficient varying out to some outer boundary r, here taken to be 120 (130) AU, and the solar wind speed V, = const = 400 km·s$^{-1}$. This is a one zone heliosphere first described by Parker, 1965. The second scenario is a two zone heliosphere (e.g., Jokipii, Kota and Merenyi, 1993). In this case the inner zone extends out to 90 AU, the average distance to the HTS. In this inner region V=400 km·s$^{-1}$ and the diffusion parameters K1 and K2 are determined in our approach by a fit to the cosmic ray data being compared (the Earth and V2) rather than using e.g., consensus values (Palmer, 1982) appropriate to the "local" heliosphere.

The outer zone extends from 90 AU to $\sim$120 (130) AU, the approximate distance to the heliopause (HP) or an equivalent "outer boundary" and essentially encompasses the heliosheath. In this region V is taken to be 130 km·s$^{-1}$ (from V2 measurements, Richardson, et al., 2008) and the diffusion parameters are K1H and K2H, which are different from those in the inner heliosphere, and again determined by the cosmic ray intensity changes at V1. The distance to the HP and the source spectrum are important in this calculation.

For the LIS proton spectrum we use the recent spectrum of Webber and Higbie, 2009. This spectrum can be approximated to an accuracy $\sim$few % for energies above $\sim$100 MeV/nuc as

$$\text{Proton FLIS} = (18.9/T^{2.79})/ (1+6.75/T^{1.22}+1.30/T^{2.80}+0.0067/T^{4.32})$$

where T is in GeV/nuc. At the average energy of 200 MeV, this equation gives an input intensity of 9.8 ±0.5 p/m$^2$·sr·s·MeV/nuc at the boundary at 120 (130) AU. The V1 intensity (at 114 AU) measured at 2010.5 is 7.6 in the same units, about 25% lower than the IS intensity. The intensity at V2 at the same equivalent time (+0.21 year) is 6.5 and the intensity at the Earth $\sim$0.89 year earlier is 2.3 in the same units.

Consider a simple heliosphere with a single boundary at 120 or 130 AU. The 1$^{st}$ step in this approach is to fit the measured intensity of 2.3 at the Earth at 2009.6. For K2=0 (no radial dependence of K) this requires values of K1 = 160 (175), respectively, for the two boundary locations. These values for K1 correspond, for each boundary location, to a modulation

potential = 250 MV in the equivalent force field approximation where the modulation potential is defined as

$$\phi = \int_1^{R_B} \frac{V \, dr}{3K1}$$

(see Caballero-Lopez and Moraal, 2004).

This modulation potential is ~4% lower than that we have recently obtained for He nuclei also at ~200 MeV/nuc (Webber, et al., 2011). This difference could be related to differences in the input spectra used or differences related to the rigidity for protons (~0.6 GV) or Helium nuclei (~1.2 GV) and the rigidity dependence of K1 used in the calculations. Both the modulation potential derived for protons (or He nuclei) are much lower than the average value of ~400-500 MV observed at previous sunspot minima in the modern era from 1950 (see e.g., Webber and Higbie 2010), in keeping with the unusually high intensities observed at this time in 2009 (McDonald, Webber and Reames, 2010; Mewaldt, et al., 2010). In fact the low modulation potential that we now find using protons or Helium nuclei is very similar to the modulation potential obtained by Mewaldt, et al., 2010 using ACE measurements of C and Fe nuclei at the Earth at the same time in 2009.

For the values of K1 which fit the data at the Earth between 2005 and 2010, the calculated intensities at V1 and at V2 do not provide a good fit to the data lines Figure 4A for a 1 zone model. If the value of K is assumed to increase with r rather than be a constant, for example, if K~r, the fit to the data lines in Figure 4A is still unsatisfactory. So it is clear that a simple one zone heliosphere cannot accurately determine the intensities simultaneously observed at V1, V2 and the Earth.

For a two zone model based on an inner heliosphere inside the HTS and an outer heliosphere (the heliosheath) between the HTS and the HP with the inner heliosphere boundary at the HTS (taken here to be at 90 AU) and the HP at 120 (130) AU, we find that, for values of the HP = 120 (130) AU, the values of K1=175 (max) and 42 (min) and K2 = 0 in the inner heliosphere and values of K1H between 18 (30) (max) and 10 (24) (min), and K2H=0 with V = 0.33 in the heliosheath as obtained for He nuclei (Webber, et al., 2011) also fit the data at the Earth and at V1 for protons in Figure 4A. These calculated lines for V1 and V2 are shown along with the data from Figure 4A in Figures 4B and 4C for $R_B$=120 and 130 AU respectively. The vertical distance between the V1 and V2 lines provides a continuous measure of the effective radial intensity gradient between these two spacecraft. The predicted V1 "line" for $R_A = 120$ AU lies an average of 2% below the data and none of the smoothed data points lie more than ±10% from the predicted line. The passage of transient structures, the largest of which occurs at 2006.15 at V2, modify the overall simple sphericity of the heliosphere considerably.

For $R_B=130$ AU shown in Figure 4C, the fit between data and predictions is less good. The predicted line is much flatter than the observed data. Of course the predictions are based on the parameters obtained from fitting the He data. Small (~10%) changes in these modulation parameters for the $R_B=130$ AU case will change the slope and improve the fit.

An important feature of this analysis is related to the fits to the V2 data. For V2 the fit is less good than for V1 data and the calculated intensities are 5-20% less than those observed. If the N-S asymmetry of the heliosphere, which is ~10% (see Washimi, et al., 2007; Opher, et al., 2009) is taken into account then the effective distance of V2 should be increased by about 10 AU and the calculated intensities should be increased by ~10-15%. This improves the fit between calculations and data considerably as seen by the dashed line in Figure 4B for 120 AU, where the differences between observations and predictions now range from ~0 to ~10%. For the calculations with $R_B = 130$ AU in Figure 4C the predicted intensity changes at both V1 and V2 are still less than observed.

Overall the V2 data for 150-250 MeV protons during the 2004-2010 time periods can be much better fit than the data for He nuclei of the same energy within the framework of a spherical but squashed heliosphere. This is mainly because protons during the time period from 2007.0-2009.0 have a non-zero positive radial intensity gradient between V1 and V2 in contrast to the zero gradients observed for He nuclei.

Thus, in summary, we have the situation where (1): The magnitude of the diffusion coefficient in the outer zone (heliosheath) as derived from both the proton and helium nuclei data is ~5-10 times smaller than that in the inner zone. But (2): During the intensity recovery from 2004-2010 the diffusion coefficient in the inner zone increases by a factor ~4 whereas in the outer zone this increase is only a factor ~1.80 (1.25). (3): The intensity differences between V1 and V2 protons can be reasonably well fit in a squashed heliosphere with $R_B = 120$ AU and somewhat less well for $R_B = 130$ AU.

## Summary and Conclusions

The recovery of the intensity of ~150-250 MeV/nuc cosmic ray protons has been followed between 2004-2010 at the Earth and also at V1 and V2 in the outer heliosphere and in the case of V1, beyond the HTS. The correlation of the intensity changes at the Earth and V1 in the outer heliosphere (correlation coefficient =0.936), ~100 AU apart, is remarkable after accounting for a time delay ~0.9 year due to the solar wind propagation. The relative intensities at V1, V2 and at the Earth as well as the slope of the regression lines between the measurements place limits on the amount of solar modulation in the inner and outer heliosphere. It is found that the data at the Earth and at V1 can be reproduced by a simple two zone heliosphere where the intensity changes are due to changes in the cosmic ray diffusion coefficient K in each zone. In the inner zone, out to the HTS assumed to be at 90 AU, the value of K is quite large (see Figure 5) but varies by a factor ~4 from the minimum to maximum modulation in this part of the solar 11-year cycle. In the outer zone from ~90-120 (130) AU, essentially in the heliosheath, the value of the diffusion coefficient is much smaller, by a factor ~5-10 and varies by a factor ~1.80 (1.25) from minimum to maximum modulation. These are the same variations of K used to describe the helium nuclei temporal variations at the same energy by Webber, et al., 2011.

In effect the heliosheath appears to be a very turbulent, diffusive region, acting much like a diffusive barrier to these lower energy cosmic rays because of the small value of the diffusion coefficient, in spite of the slower solar wind which tends to reduce the effect of adiabatic energy loss.

Although the V1 proton data is fit to a level ~±5% over the entire time period from 2005-2010 for boundaries between 120-130 AU, the V2 proton data is not as well fit with a simple spherically symmetric heliosphere, with the predictions typically ~5-20% less than the data. If the heliosphere in the V2 direction is assumed to be flattened with an asymmetry as determined by Washimi, et al., 2007 (see also Opher, et al., 2009), then the model fit to the V2 data is generally better (the differences are now ~10% or less), and can be made still better with slight changes in the time dependence of the modulation parameters that make them different than those that were used for He.

The details of the fit to the data beyond the HTS depend on the values of the local interstellar spectrum (LIS) used as an input to the modulation calculation and also the location of the HP or boundary to the modulation region. For the estimated IS intensity of protons used in this paper the data can be fit for HP distances in the range of 120-130 AU. This heliosheath region and the interstellar proton spectrum itself will be mapped in more detail as V1 continues to move outward in the heliosphere and the intensity continues to increase towards the LIS value.

**Acknowledgements:**  The authors wish to thank the Voyager team (E.C. Stone, P.I.) and the ACE team (R.A. Mewaldt, P.I.) for making their data available on their web-sites, http://voyager.gsfc.nasa.gov and http://www.srl.caltech.edu/ACE/.

**References**

Caballero-Lopez, R.A. and H. Moraal (2004), Limitations of the force field equation to describe cosmic ray modulation, J. Geophys. Res., <u>109</u>, A01101, doi:10.1029/2003JA010098

Gleeson, L.J. and I.A. Urch, (1971), Energy losses and the modulation of galactic cosmic rays, Astrophys. Space Sci., <u>11</u>, 288-308

Jokipii, J.R., J. Kota and Merenyi, (1993), The gradient of galactic cosmic rays at the solar wind termination shock, Ap. J., <u>405</u>, 753-786

McDonald, F.B., W.R. Webber, E.C. Stone, A.C. Cummings, B.C. Heikkila and N. Lal, (2006), Voyager observations of galactic and anomalous cosmic rays in the Heliosheath, AIP Conf. Proc., <u>858</u>, 79-85, doi:10.106311.2359309

McDonald, F.B., W.R. Webber and D.V. Reames, (2010), Unusual time histories of galactic and anomalous cosmic rays over the deep solar minimum of cycle 23/24, Geophys. Res. Lett., <u>37</u>, L18101, doi:10.1029/2010GL044218

Mewaldt, R.A., R. Leske and K. Lave, (2009), Cosmic ray Fe intensity reaches record levels in 2008-2009, ACE News #122, http://www.srl.caltech.edu/ACE/ACENews/ACENews122. html

Mewaldt, R.A., et al., (2010) Record-setting cosmic ray intensities in 2009 and 2010, Ap.J. Lett., <u>723</u>, L1-L6

Opher, M., J.D. Richardson, G. Toth and T.I. Gombosi, (2009), Confronting Observations and Modeling: The role of the Interstellar Magnetic Field in Voyager 1 and 2 Asymmetries, Space Sci. Rev., <u>143</u>, 48-55

Palmer, I.D. (1982), Transport coefficient of low-energy cosmic rays in the interplanetary space, Rev. Geophys. Space Phys., <u>20</u>, 335

Parker, E.N., (1963), Interplanetary dynamical process, New York, Interscience

Reinecke, J.P.L., H. Moraal and F. B. McDonald, (1993), The cosmic radiation in the heliosphere at successive solar minima: Steady state no-drift solutions of the transport equations, J. Geophys. Res., <u>98</u>, (A6), 9417-9431

Richardson, J.D., J.C. Kasper, C. Wang, J.W. Belcher and A.J. Lazarus, (2008), Cool heliosheath plasma and deceleration of the upstream solar wind at the termination shock, Nature, <u>454</u>, 63-66, doi:10.1038/nature07024

Washimi, H., G.R. Zank, Q. Ha, T. Tanaka and K. Munakata, (2007), A forecast of the heliospheric termination shock position by three dimensional MHD stimulation, Ap.J., <u>670</u>, L139-L142

Washimi, H., G.R. Zank, Q. Ha, T. Tanaka and K. Munakata, (2010), Realistic and time-Varying Outer Heliospheric Modeling by Three-Dimensional MHD Simulation, Ap.J., (in press)

Webber, W.R. and D.I. Intriligator, (2011), Voyagers 1 and 2 in a Shrunken and Squashed Heliosphere, JGR, <u>116</u>, A06105, doi:10.1029/2011JA016478

Webber, W.R., F.B. McDonald, P.R. Higbie and B. Heikkila, 2011, Recovery of 150-250 MeV/nuc cosmic ray helium nuclei intensities between 200-2010 near the Earth, at Voyager 2 and at Voyager 1 in the heliosheath – A two zone heliosphere,

Webber, W. R. and P. R. Higbie, (2009), Galactic propagation of cosmic ray nuclei in a model with an increasing diffusion coefficient at low rigidities: A comparison of the new interstellar spectra with Voyager data in the outer heliosphere, J. Geophys. Res., <u>114</u>, A02103, doi:10.1029/2008JA013689.

**Figure Captions**

**Figure 1:** 5 x 26-day running average of V1, V2 and IMP/ACE 150-250 MeV/nuc proton intensities from 1998 to 2010.5. The Earth data is delayed by 0.89 year to account for inner-outer heliosphere delay in modulation due to solar wind propagation time. The blue data points are from the IMP-8 ACE correlation.

**Figure 2:** 5 x 26 day running average of the V1 and V2 (in red) to Earth intensity ratio of 150-250 MeV/nuc protons from 2004 to 2010.5 (Earth data delayed by 0.89 year).

**Figure 3:** The V1 data from 2004.8 in Figure 1 superimposed on the data at the Earth delayed by 0.89 year (with different intensity scales on the left and right axis). This figure shows the level of correlation between intensity changes at the Earth and in the outer heliosphere during this time period.

**Figure 4A:** Regression plot of the intensities from 2004.8 to 2010.5 at V1 (black) and V2 (red) vs. the intensities at the Earth delayed by 0.89 year. Both vertical and horizontal axis in figures 4A, 4B and 4C are in P/m$^2$·s·sr·MeV.

**Figure 4B:** The V1 (black) and V2 (red) data points from Figure 4A superimposed on the model predictions, ($R_B$ = 120 AU), K1H = 18 (max) to 10 (min). The effect of a general heliospheric radial N-S asymmetry ~10% near the HTS on the predictions for V2 is shown as a dashed line.

**Figure 4C:** Same as Figure 4C but with $R_B$ = 130 AU and K1H changing from 30 (max) to 24 (min).

**Figure 5:** Values of K1 and K1 H used in the two zone modulation model. Black lines labeled 2010 and 2005 show the range of values of K in the inner heliosphere and also in the heliosheath necessary to reproduce the proton intensity changes observed between 2004 and 2010.5 at the Earth and at V1 and V2. The solid points at 1 GV indicate the maximum and minimum values of K1 and K1H at that rigidity.

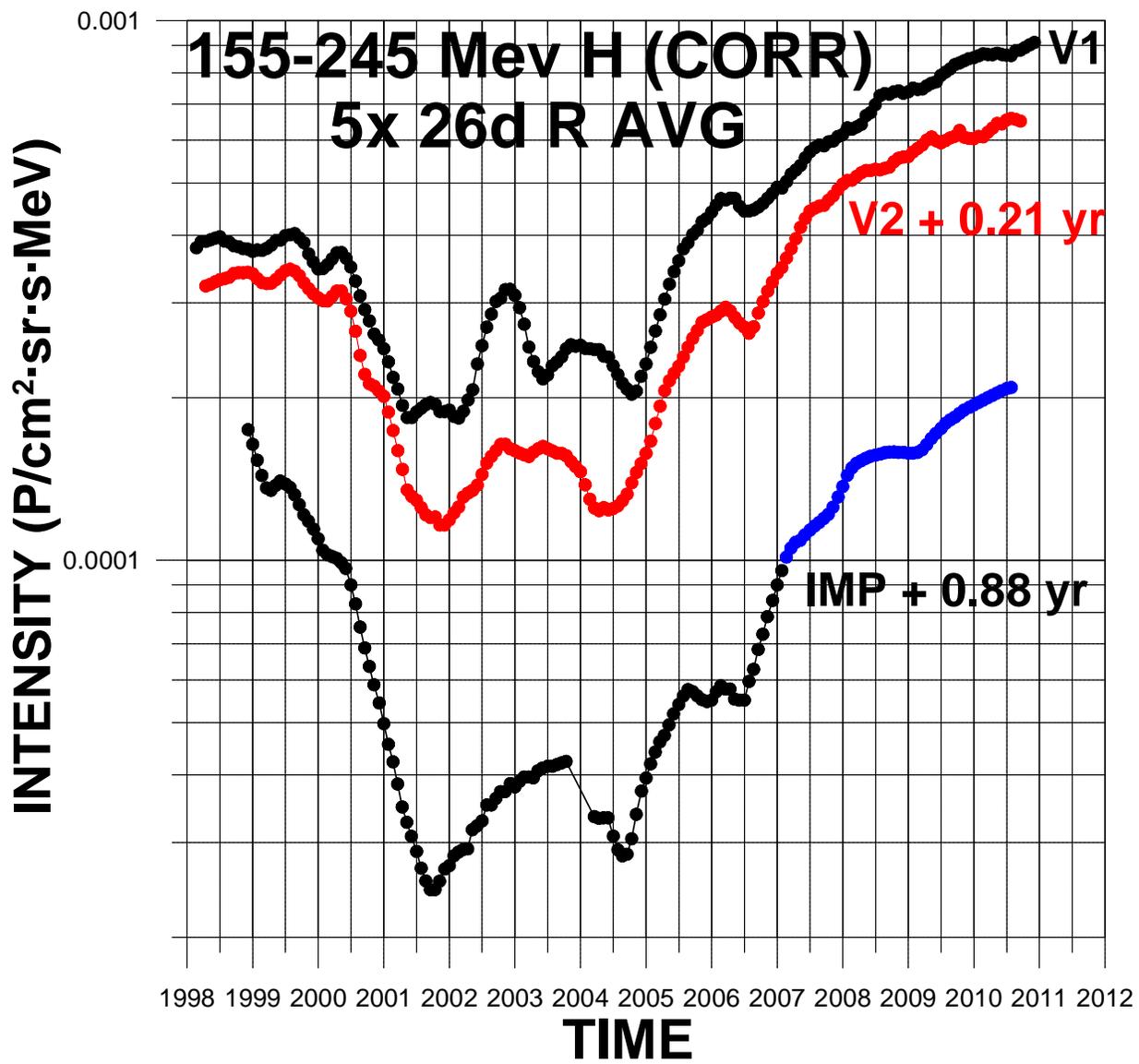

FIGURE 1

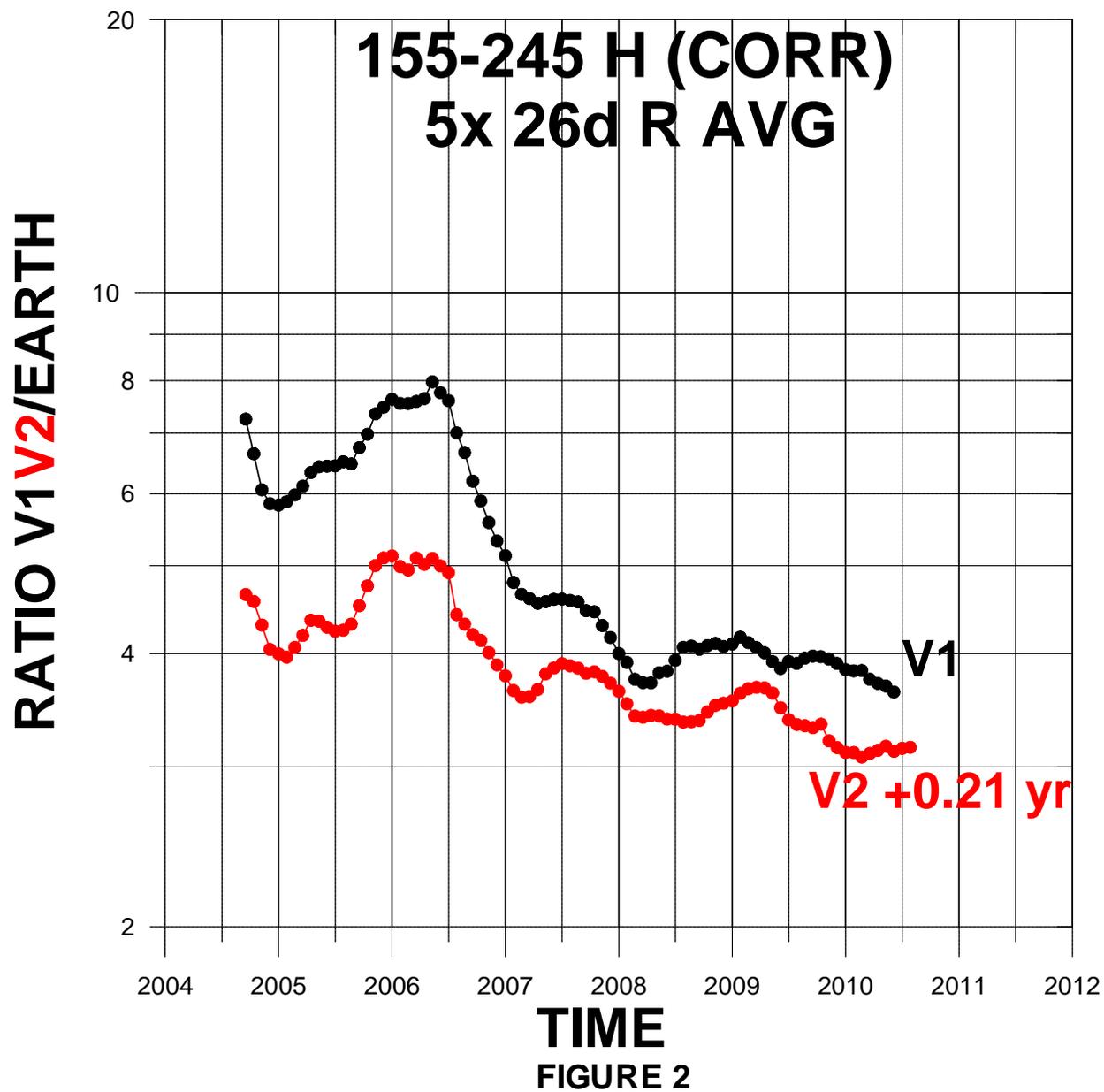

FIGURE 2

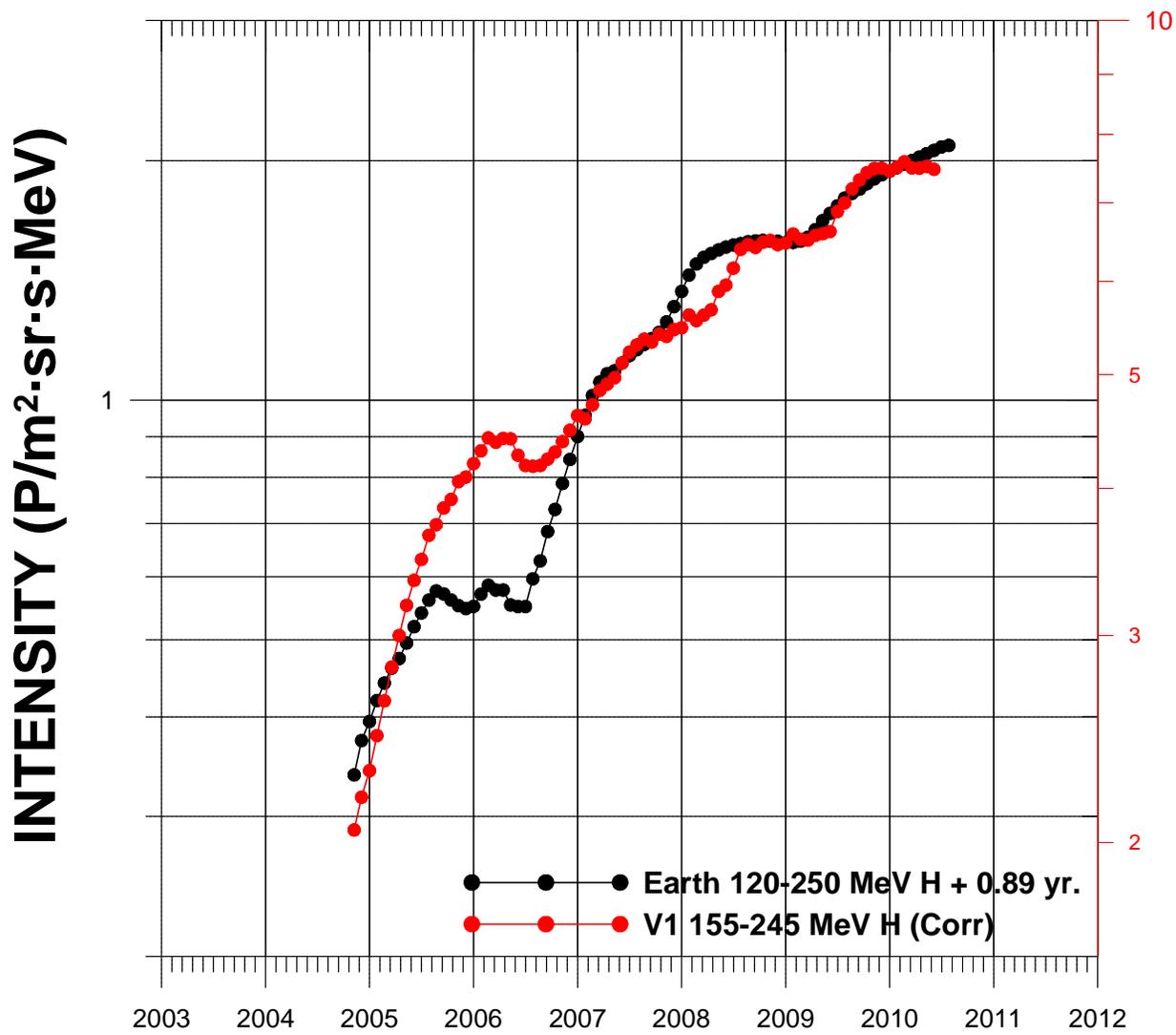

**TIME**

**FIGURE 3**

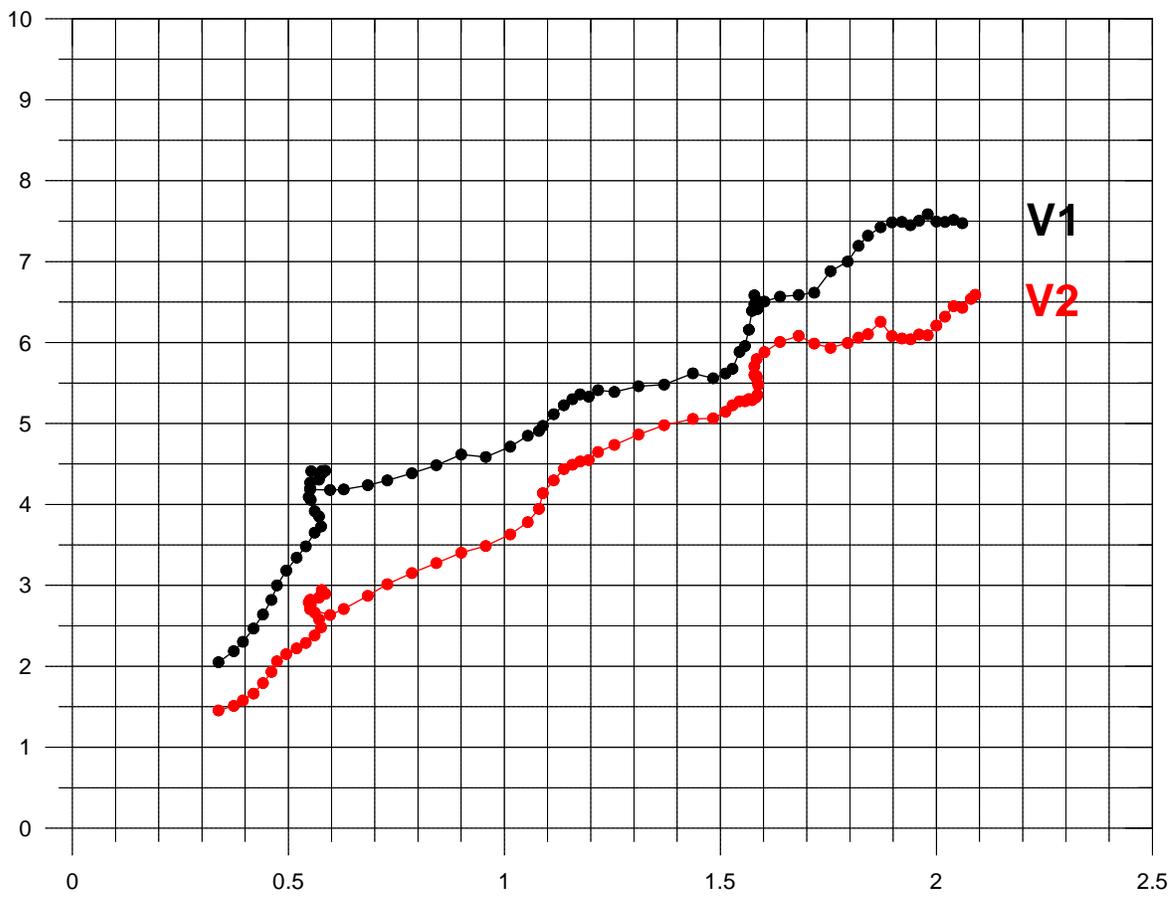

**FIGURE 4A**

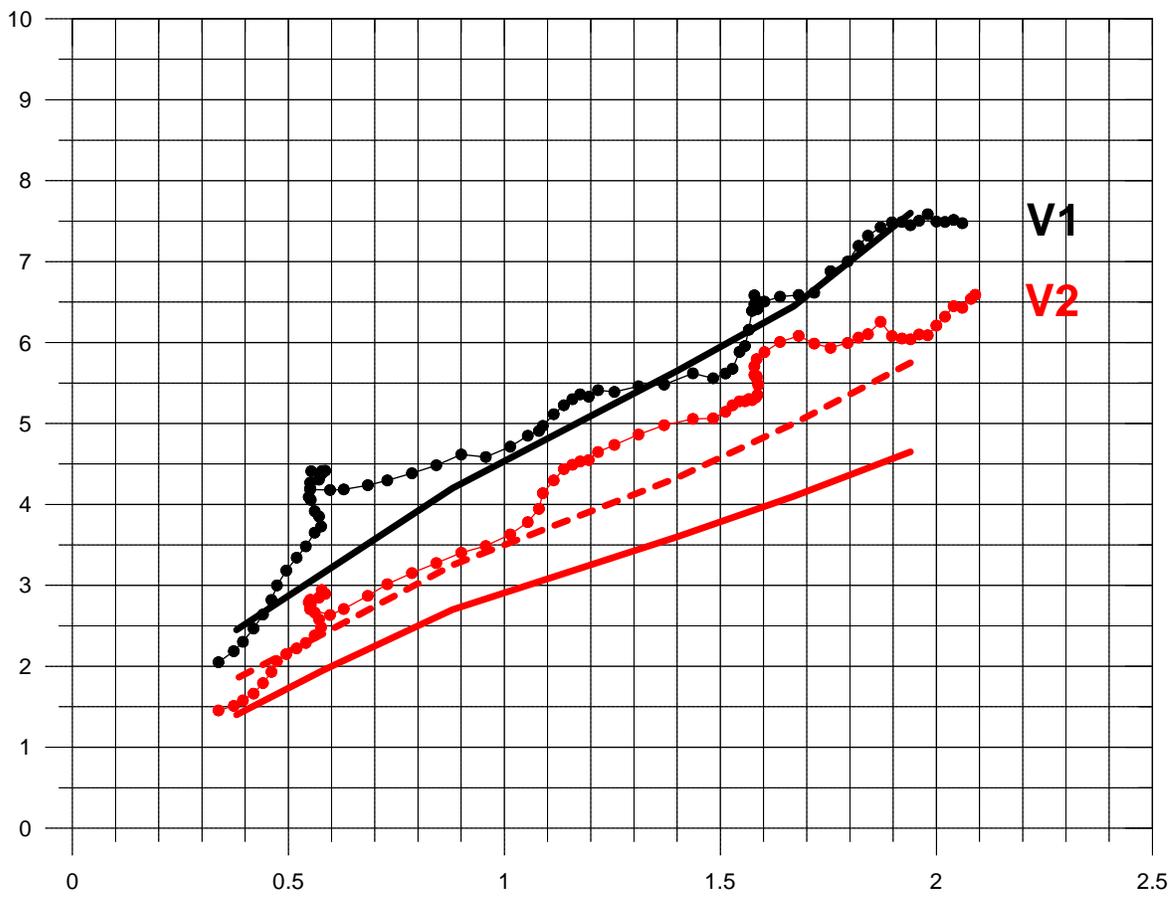

**FIGURE 4B**

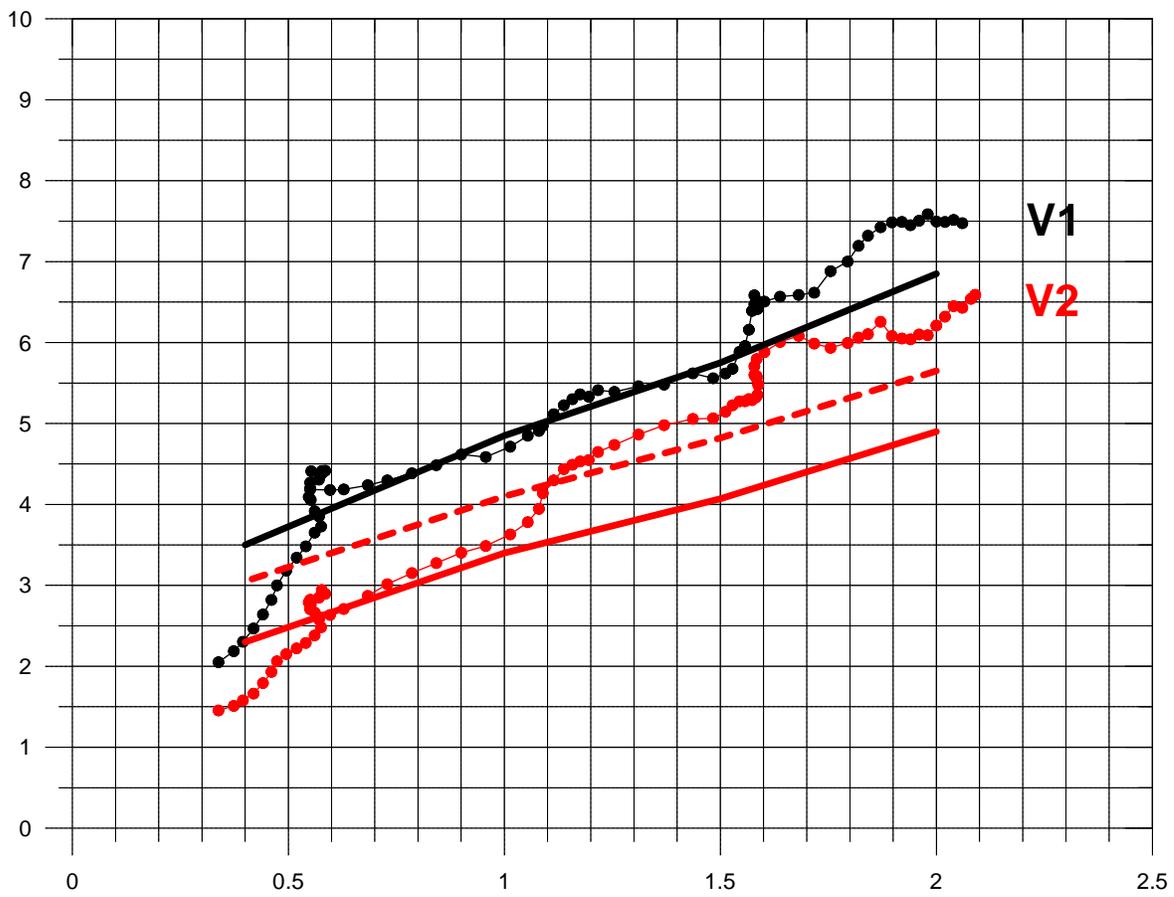

FIGURE 4C

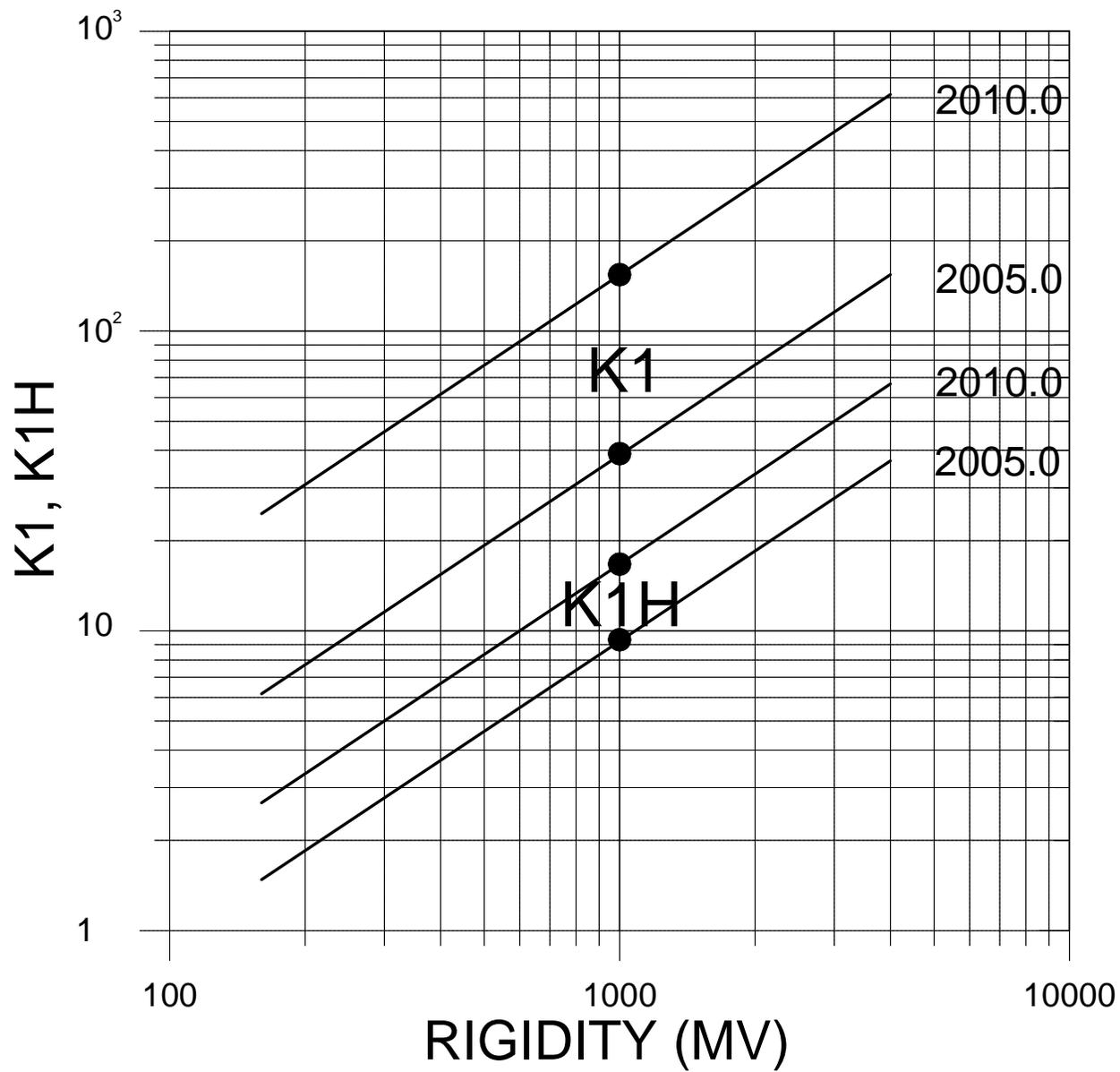

FIGURE 5